# Sea Level Enigmatic Rising
## New Perspectives from an Expanding Globe

by
**Giancarlo Scalera**
*INGV – Istituto Nazionale di Geofisica e Vulcanologia*
*Via di Vigna Murata 605 – 00143 Roma, Italy*
*(retired)*

**Abstract:** In the expanding Earth framework it is possible to find additional phenomena that could contribute in a proper way to the water balance and general tectonic eustatism involved in the sea lever rising. Recent compilations seems to leave unexplained up to 12 cm/century of sea rising, and possible solutions invoking a polar ice shells melting near to the upper limit of the error bars reveal in conflict with the consequent expected decreasing of the Earth angular velocity. It is shown that taking into account possible effects of an expanding Earth, the problem can be initiated towards an appropriate solution, at least as regards the just orders of magnitude. Major effects on sea-level could come from ongoing relaxation of curvature variations that are peculiar for an expanding globe.

**1 – Introduction**

Because a substantial expansion of the globe must have consequences on the evolution of the oceanic basins and their contents, it is obvious to scrutinize if the results of recent researches about sea level rise and their unsolved problems can be explained in the expanding Earth framework.

The expanding Earth is a concept that should be considered the next step toward the true mobilism and a more complete evolutionistic view of Earth's geologic processes – eventually to be an interpretative key also in planetology. While plate tectonics conserves an hidden heritage of old fixist and contractionist ideas – especially in its orogenetic compressional corollary – that does not conflict with previous Anglophon Academic tradition (Brouwer, 1981; Scalera, 2012a), instead, expanding Earth is providing new harmonic explanations for a number of phenomena that were thinked splitted in two separated processes, like the mountain ranges and the mid-oceanic ridges, the secular Polar Motion (PM) and the True Polar Wander (TPW), and so on (Besse & Courtillot, 1991, 2002; Scalera, 2003, 2006, 2012a, 2013).

What become important in this conception is the emissive activity of the planet, which shows in volcanic eruptions (Scalera, 2013), rising of megadykes (Ollier, 2003, 2012; Scalera, 2012d), reservoirs of methane gas (Katz et al., 2008; Scalera, 2012b; among others) etc., besides an isostatic mechanism of mountain building, without a large scale subduction (hundreds of km of underthrust) but allowing for the overthrusting or underthrusting of few tens of km, often observed on the field (Scalera, 2010, 2012d).

In this context of more radical evolution of the planet, even the change of the surface of the ocean floor (increasing, at least from the beginning of the Mesozoic), the change in their water content, and changes – both regional and

global – in the shape of the Earth, should be considered in analyzing the data of sea level.

## 2 - Problems in sea level rise

It is a long time that the variation of the sea level has attracted the interest of geosciences community (e.g., Fairbridge, 1961; Carey, 1981; Pugh, 1990; Hallam, 1992; Aaron et al. 2010; Lambeck et al., 2010; De Santis, 2012; among many others), but recent discussions about the rising of the sea level (Church, 2001, 2006; Munk, 2002; Douglas & Peltier, 2002; Cazenave & Nerem, 2004; Miller & Douglas, 2004) have put in evidence the existence of a probable discrepancy between the estimated and measured sea level secular rise after the end of the last *little ice age* early in the 19th century. Global sea level rise (GSLR) and its causes are then subjects of an intense controversy called the *attribution problem* (Miller & Douglas, 2004).

The total sea level rise $\zeta_T(t)$ (referred to the crust, which in turn can move vertically) is splitted in two parts, the first linked to variation of the amount of mass of ocean water and the second to variation of ocean volume (without variation of mass):

$$\zeta_T(t) = \zeta_e(t) + \zeta_s(t) . \qquad (1)$$

The first term $\zeta_e(t)$ is called *eustatic* and is generally credited to be linked to:

i) growth or melting of glaciers.

ii) deposition of new snow cover on glaciers or creation of new glaciers.

iii) creation or growing, and depletion or exhaustion of any kind of non-marine underground water reservoir and aquifer.

The second term $\zeta_s(t)$ is called *steric* and ascribed to the thermal expansion of the sea water.

Recent estimates of the global sea level rise point to values of 1.5-2.0 mm/y, a rate that should be distributed on the first and second term of equation (1). The estimates for possible maximum values of the two terms coming from glaciers melting and thermal expansion of the saline oceanic water – as consequence of green-house effect – are both many times inferior to the needed (Munk, 2002; Miller & Douglas, 2004). The needed supplemental heat storage in the oceans should be $10^{24}$J per century, while an estimate of only $2.0 \cdot 10^{23}$J per century is the maximum value allowed by data and theory, and the difference between the measured (21 cm) and the maximum contribution eustatic (6 cm) and steric (3 cm; due essentially to greenhouse) for the last century is

$$\zeta_{measured}(2000) - \zeta_e - \zeta_{greenhouse} =$$

$$= 21 - 6 - 3 = 12 \text{ cm}. \qquad (2)$$

Munk (2002) states that in these last few years – as soon as oceanographic data in heat storage have improved – the discrepancy is become more clear. It should be also recalled the ongoing discussion between IPCC (Intergovernmental Panel on Climate Change; Solomon et al., 2007; Aarup et al., 2010; Church et al., 2013) and NIPCC (Nongovernmental International Panel on Climate Change; Idso et al., 2013), which has profound implications for the reliability of the estimates of global warming and the consequent thermal expansion of the oceans.

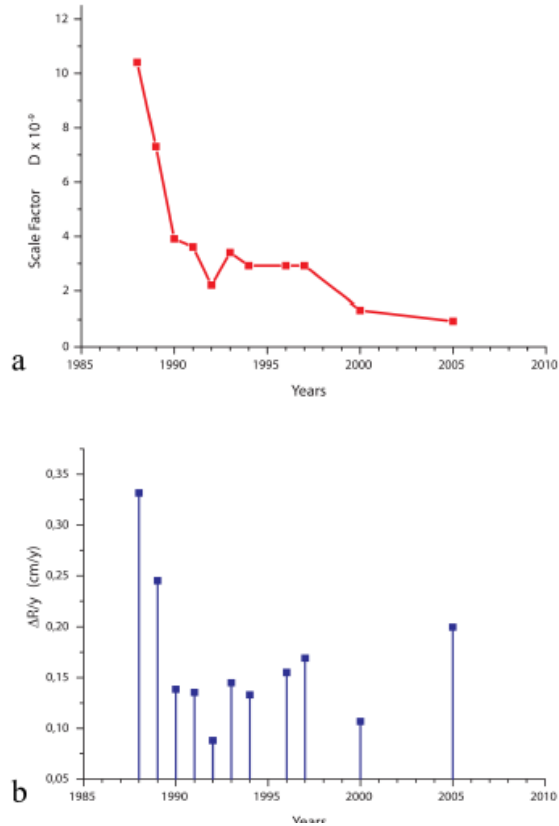

Fig. 1. – a) Values of the *scale factor D* – namely the size of the geodetic network – at different years, with respect to the ITRF-2008 (data published in the IERS Technical Notes and Annual Reports). – b) The values of the radius variations annual rate $\Delta R=y$, averaged on the time lapses from the indicated year to 2008. With the exception of the probably spurious values of 1988 and 1989, the series seems to indicate a value ≈0.15 cm/y. This means a total expansion of ≈3.0 cm on 20 years. It is worthy to note that this rate is in the order of magnitude of the discrepancy shown in equation (2) of this paper.

## 3 - Global processes changing sea level

The phenomena generally taken into account as influencing eustatic sea level at different time scales are (Conrad, 2013):

1) Elastic deformation of the crust – in response to glaciers melting, new glaciers creation, variation of the crustal load of any origin (no time delay).

2) Viscous deformation of the crust – in response to deglaciations, glaciations or others long term load variations (with time periods of $10^3$–$10^5$ yr).

3) Ocean ridge volume – in response to new segments of ridges creation, and to variation of tectonic activity.

4) Marine sediments variable accumulation or destruction – in response to climatic variation of erosion rates, or recycling of sediments into the mantle by variable hypothetical subduction rates.

5) Seafloor volcanism – in response to variation of conditions of the Earth's interior, also large submarine plateau or igneous provinces can be created.

6) Total oceanic basins area – in response to coastal erosion, inner continental rifting, multicausal continental growth at their margins, final filling or uplift of inner seas involved in hypothetical continental collisions, desiccation of large basins (like Messinian salinity crisis of Mediterranean).

7) Total continental area – in response to coastal erosion, inner rifting, multicausal continental growth at their margins.

8) Water exchange – in response to mantle water outpouring, atmospheric dissociation and loss, hypothetical re-hydratation of mantle by subduction.

## 4 - Expanding Earth global processes relevant to sea level change

In Earth expansion conception the growing of the globe and its different expressions on oceanic and continental areas could affect the relative level of waters and lands.

Efforts to directly reveal by geodetic methods the expansion rate of ≈ 1.5 cm/y indicated by variable radius paleogeographic reconstructions (from Triassic to Recent; Maxlow, 2001; Scalera, 2001, 2003) have been fruitless – the found submillimetric rates are lesser than error bars (Heki et al., 1989; Kostelecký & Zeman A. 2000; Gerasimenko, 2003, Shen et al., 2011; Wu et al., 2011; Devoti et al., 2012; Sarti, 2012; Scalera, 2012c). An indirect indication can be found in the *scale factor* D of the global geodetic network (Scalera, 2012c), pointing to an increasing radius value of ≈3.0 cm on twenty years (≈1.5 mm/y) (Fig. 1ab).

This lower than expected value can be put in relation to the global *tectonic activity* map (Müller et al., 1997; McElhinny & McFadden, 2000). The *Half Spreading Map of the Oceans* (Fig. 2) presents clear minima at the same ages – Recent, Cretaceous-Cenozoic boundary, Jurassic-Cretaceous boundary – in all the oceans, constituting an independent support both to the assumption that global expansion is in a *stasis* state today, and to reliability of scale factor indication. It must be stressed that the projection of the rate indicated by the scale factor D to the secular rate – ≈15 cm/century – is in the order of magnitude of the discrepancy shown in equation (2) of this paper; which means nothing but that the matter needs to be investigated.

*Curvature change and its dissipation*
Thick continental region lithosphere can be in first approximation supposed of constant area, but must tend to conserve the old higher curvature during the increase of the Earth's radius – that is to say decreasing curvature. The mechanism is illustrated by Hilgenberg (1933) and after him many others have used this idea to explain orogenetic phenomena (Matschinski, 1954; Rickard, 1969; Cwojdziński, 1991, 2003).

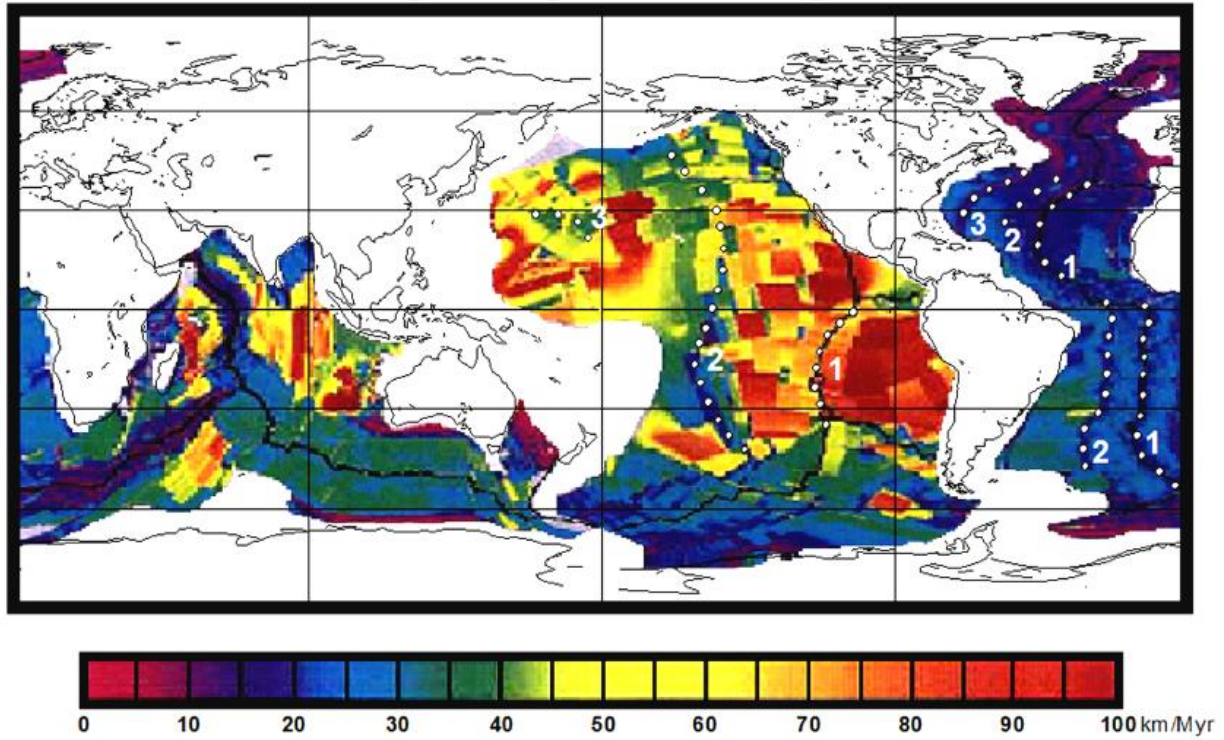

**Fig. 2.** In the *Half Spreading Map of the Oceans* (Müller et al., 1997; McElhinny & McFadden, 2000; Müller et al., 2008) at least three periods of slowdown of sea-floor expansion are present: the Recent, the Cretaceous-Cenozoic boundary and Jurassic-Cretaceous boundary, indicated by white circles and the numbers 1, 2 and 3 respectively.

In Fig. 3a is shown the ideal perfectly rigid crust and lithosphere – a fragment of a quarter of a great circle – transferred from an initial sphere of radius $R_i$ to a series of spheres with

$$R_n > R_{n-1} > \ldots > R_k > R_i.$$

The maximum eight $h_m$ of this ideally non-dissipating bulge follows the function of $R$ shown in Fig.4a approaching to the limit:

$$h_m = \lim_{R \to \infty} h(R) = R_i \cdot (1 - \cos(\theta/2)), \qquad (3)$$

and its increments with respect to the preceding step of the radius variation ($\Delta R = const$) follows the function in Fig. 4b annulling asymptotically to $R=\infty$. It is also easy to see that in this particular case with initially $\theta=90$ the rate $\Delta h(R)/\Delta R = 1/\sqrt{2} \approx 0.71$ at beginning of expansion, decreasing toward 0 at $R=\infty$. The meaning of this is that the response of the uplift to an increasing of radius is in the same order of magnitude of the $\Delta R$.

For the real Earth, assuming an initial radius $R_i = 3600$ km, the maximum eight of the ideal bulge is $h_m \approx 1000$ km, which plausibility is obviously poor and needs to be verified.

The same phenomenon can influence the sea level. Albeit erosion can mitigate this effect, the rising of the central continental regions – enhanced going from coastal edges towards their central regions – must undergo isostatic compensation. Continents should then show a behavior like the glacial isostatic adjustment, with a coastal subsidence proportional to their inner uplift. An evidence that this could be the case is the existence of a perimetral submerged bulge all along continental coasts (Catuneanu, 2004; Allen & Allen, 2005). The order of magnitude – fraction of mm/yr – is just as needed to close the *attribution problem*, but this physical process evolves in strict parallelism to the erosive one, which enhances its effect.

*Erosion of the rising land bulges*

Besides the uplift of the land bulges caused by decreasing curvature of the globe, the erosion acts as a limiting process that forbids $h(R)$ both to approach the limit indicated in equation (3) and the surface $S_1$ in Fig. 3b. We can suppose that the amount of mass between the surfaces $S_1$ and $S_2$ in Fig. 3b is the maximum amount of material to be discharged to the ocean basins.

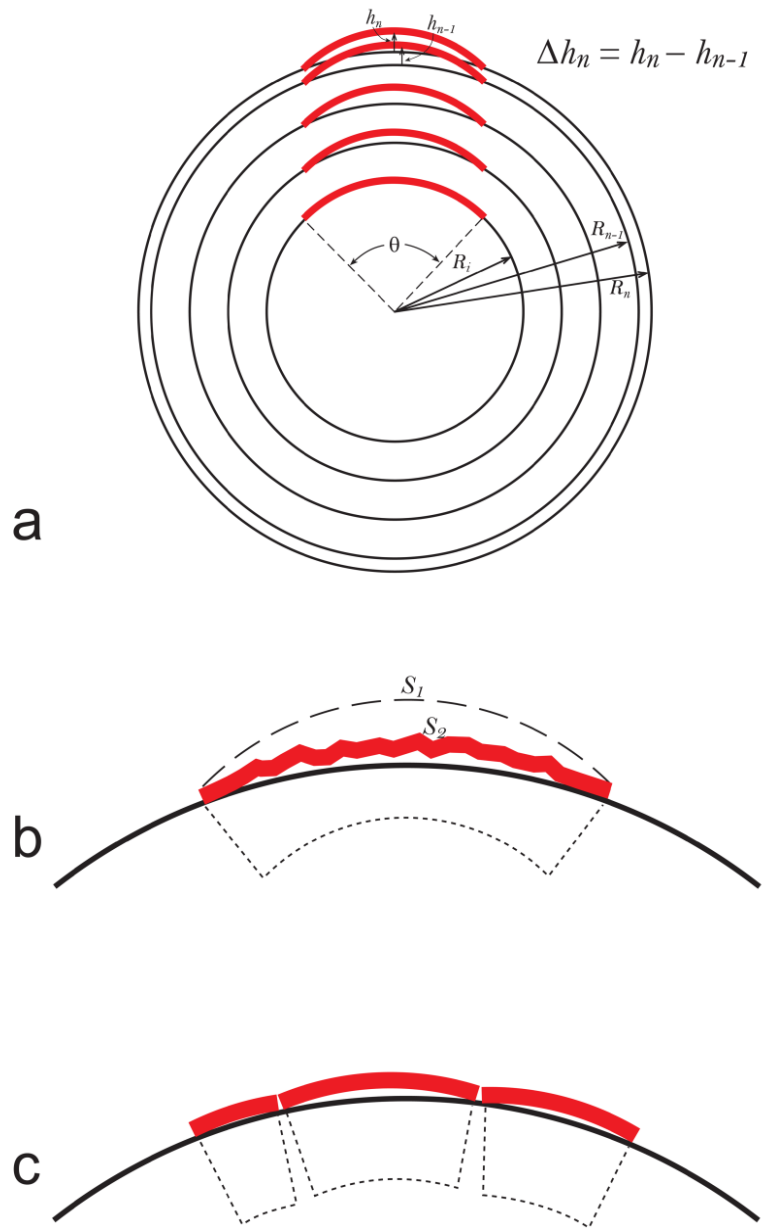

Fig. 3. – a) – A portion of a sphere which measures a quarter of circumference is ideally transferred without deformation from the sphere of radius $R_i$ to a series of spheres of radiuses $R_k > R_i$. The height of the protuberance $h(R)$ increases with the radius following the trend of the function in Fig. 4a. The maximum value is reached asymptotically at $R = \infty$. At subsequent increases in radius $\Delta R = const$, the increase in height of the protuberance $\Delta h$ decreases according to the function in Fig. 4b. In the real case, the erosion does not allow to reach a situation like a), but it will maintain a curvature as in b) similar the one observable today. The mass of material between the two surfaces $S_1$ and $S_2$ must be discharged into the oceans during the geologic time. Additionally, the bulge may could be fragmented as in c), an occurrence that contributes to lowering $h$. A superimposition of the states b) and c) may be more near to the real case, in a unknown proportion.

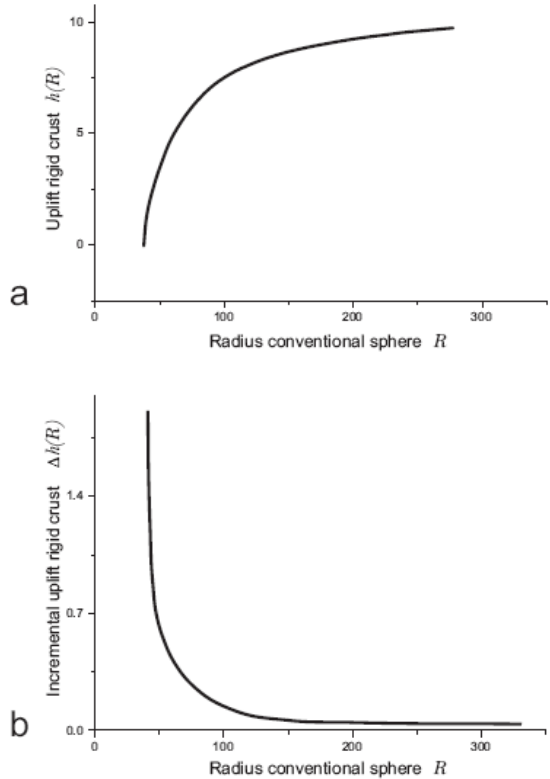

**Fig. 4.** – a) The function $h(R)$ approach asymptotically the maximum value $h_{max} = 1 - \cos(\vartheta/2)$. – b) The increase in height of the protuberance $\Delta h(R)$ decreases asymptotically to 0 when $R$ approach $\infty$.

But considering the possibility of an anelastic fracture of the lithosphere (Fig. 3c) the real amount should be between the maximum and an unknown minimum – but presumably a little fraction of the maximum.

It is convenient to evaluate the orders of magnitude involved in the process. A rough estimate for a Triassic Earth covered of six circular lithospheric shields, a quarter of great circle in diameter, provides for the volume of eroded debris to be transferred to the oceanic basins during the expansion of the Earth from a radius R = 3600 km to the present radius R=6370 km of:

$V_{max} \approx 40 \cdot 10^9$ km$^3$ .

Which is too large, if compared to the estimated present volume of oceanic waters of:

$V_{ocean} \approx 1.34 \cdot 10^9$ km$^3$ .

This discrepancy make clear that it is ever more realistic to assume as working hypothesis the situation shown in Fig. 3c.

A series of estimate of the global suspended sediment flux to the oceans (Milliman & Meade, 1983; Milliman & Syvitski, 1992; Ludwig & Probst, 1998; Beusen et al., 2005; Walling & Webb, 1996, pag.7) has progressively lead to an amount of about $20 \cdot 10^9$ Ton/yr. Which assuming a mean density for the continental emerged crust of $\rho$=2.4 g/cm$^3$ allows to compute an annual rate of eroded volume:

$V_{eroded} \approx 8.3$ km$^3$/yr .

The surface of the continents $S_{cont}$ is $\approx$30% of the total Earth's surface area $S_E \approx 5.1 \cdot 10^5$ km$^2$.

Assuming a mean continental altitude of 0.8 km, the volume of the emerged part of the continents is

$V_{emerged} \approx 1.2 \cdot 10^8$ km$^3$ .

The comparison of $V_{emerged}$ with the annual rate of volume eroded $V_{eroded}$ allows to estimate the time $T_d$ needed to destroy the emerged lands:

$T_d = V_{emerged} / V_{eroded} \approx 14.7$ Myr .

The preceding results constitutes a paradox: the emerged portions of the continents would not survive for more than few million years – and not more than 150 Myr if more conservative assumptions about plane lowlands are adopted.

The difficulty persists also taking into account the isostatic rebound of the continental crust, which – assuming a constant radius – would have to lift toward the surface the lower crust in 150 Myr also in all the coastal regions, that is a fact unacknowledged by geology.

This problem can be successful overcome by assuming a process of continental steady emersion, enhanced toward their interior regions, caused by curvature variations – schematically shown in Fig. 3b and Fig. 3c – as effectively acting with continuity during the geologic time.

This solution allows for an emersion of deep materials in the shields regions and forbids the same uprising along the coasts.

It is also interesting to evaluate how much volume of sediments $V_{250}$ is discharged to oceans – if the actual rate of global erosion is supposed constant – in 250 Myr, namely the lapse of time of doubling of the Earth's radius:

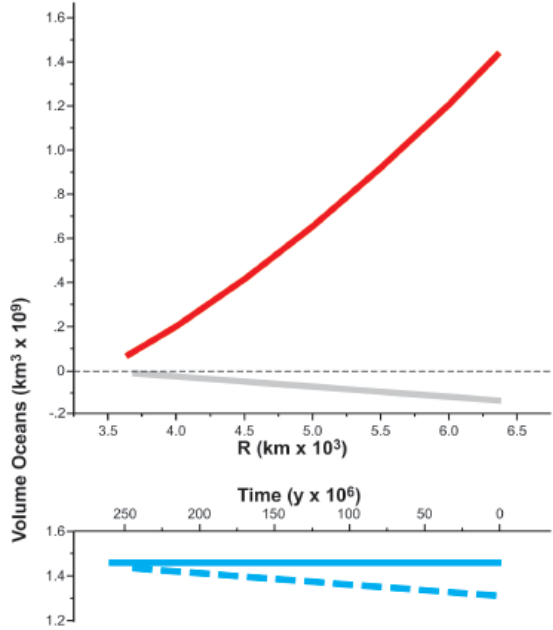

**Fig. 5.** The oceanic water volume during geologic time. The red bold line represents the possible increase of the global oceanic volume as soon as the Earth's radius increase from 3700 km in Triassic to the Recent value. In this conception no need exists to be in agreement with the loss of oceanic waters – their illogical reintroduction into the mantle – along the subduction zones, with the rate value taken from the next Fig. 6 and represented by the grey bold line. In the plate tectonics framework the sea waters volume was initially conceived as roughly constant – represented by the bold blue line, but the blue dotted line is obtained by applying the decrement rate of the grey line.

$$V_{250} = V_{eroded} \times 250 \text{ Myr} \approx 2.1 \cdot 10^9 \text{ km}^3,$$

which, reasonably, is

$$V_{250} \approx (1/20) \cdot V_{max}, \text{ and}$$
$$h_{250} \approx (1/20) \cdot h_m,$$

providing a rough evaluation of the real annual bulging $h_{annual}$ due to curvature decreasing.

$$h_{annual} \approx (h_{250}/250 \text{ Myr}) \approx$$
$$\approx 0.2 \text{ mm/yr} = 2.0 \text{ cm/century}$$

again in the needed order of magnitude to be useful to close equation (2).

*The expanding global ocean basin*

Adopting the expanding Earth framework, the surface of the oceans must be considered not as a constant but as a variable. Their total surface has increased starting from Triassic, from a small amount to the actual one. Several clues have been found (Dickins et al., 1992) that also the depth of the oceans has increased from few hundreds of meters – essentially as shallow epicontinental seas – to their actual mean depth of 4000 m – but deeper than 10000 m in the trench zones.

This expansion, beside the assumption of a rough constancy of the waters volume, has led many expansionists (e.g. Egyed, 1956) to consider that a regression of the seas should be observable in the time series of maps of paleogeographical atlases, and is a proof of the theory. The progressive increase of dry-land has been confirmed by Hallam (1992), Smith et al. (1994; in the 70-250 Ma time interval), and others, also if their preferred explanations does not involve Earth's expansion.

The deposition of the eroded continental debris on the seafloor is equivalent to the creation of an additional layer between the two geologic layers of basalt and water.

Three mutually compensating effects – at least partially – are acting:

1) The enlarging room of the expanding basins – whether as enlarging bottom surface or as increasing of average depth; resulting in a lowering of the sea level.

2) The filling of this expanding room by continental sediment flux; resulting in a rising of the sea level.

3) The subsidence of the sea floor due to the isostatic readjustment in response to the increased sedimentary load; resulting in a lowering of the sea level.

For roughly evaluate the effect on sea level of process 1), an arbitrary assumption can be a linear increase of the Earth's radius in the time lapse Triassic-Recent – about 250 Myr.

A second arbitrary assumption is that in first approximation the surface area of the continents is a constant.

The third assumption is the constancy of the erosion products that are transported into the oceans.

The function of the increasing oceans volume from Triassic to Recent is

$$V_{oce}(R) = (S_{sphere}(R) - S_{con}) \times h_{oce}(R).$$

## 5 - Not closed water cycle: mantle water content

A further main *philosophical* difference between plate tectonics and expanding Earth is about the *not closed water cycle*. Both conceptions agree that the global water cycle is not closed, but in a substantial different sense.

Plate tectonics has to take into account the conveyor belt of subduction (Deming, 1999; Bolfan-Casanova, 2005; Ohtani, 2005), which reintroduce into the mantle more water than the outpoured one (Fig. 6; data from Ohtani, 2005). Albeit it is accepted by the current theory, this fact is disturbing the common sense and leads to the equally grotesque problem of *where to store* the subducted water inside a planet that is self-drying its surface. As it can be seen in Fig. 5, the unbalanced rate of 0.67 km³/y would lead to a loss of more than 10% of the ocean water content in 250 My, and to a complete loss in 2.5 By. This is an umpleasant and somewhat illogic geodynamic property of the Earth that appears as a planet that is reabsorbing more than it is emitting. Many people have discussed this paradox (see for example the otherwise very interesting and didactical paper of Deming, 1999; among many others) as if it does not existed, and this can be considered a paradox into paradox (perceptive – psicology of *Gestalt* – and theoretical).

The positive aspect of these investigations (Bolfan-Casanova, 2005; Ohtani, 2005; Hirschmann, 2006; Hirschmann & Kohlstedt, 2012) is that – reversing the glove – there can be better understood the *storage regions from where* the waters are uplifted towards the surface (Wyllie, 1971).

As that concerns the relevance of this unbalanced loss of water on the problems of the sea level variations, the prevailing of reintroduction of water into the mantle (Ohtani, 2005) upon the outpouring of water at arcs and midoceanic ridges can eventually deflect the bold blue line of Fig. 5, and applying the decrement rate of the grey line, the blue dotted line is obtained. The ocean volume should had decreased of ≈ 0.17 × 10⁹ km³ in 250 My, like to say a total decrease of sea level of ≈ 0.48 km, which is a negligible secular lowering of ≈ 0.02 cm/century in equation (2).

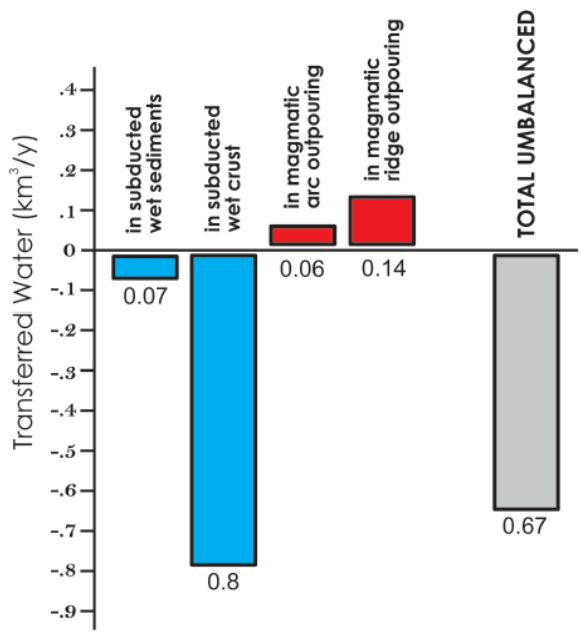

Fig. 6. The unbalanced mantle's water cycle (drawn using the data from Ohtani, 2005). The water that plate tectonics theorizes is reintroduced into the mantle (in blue) is many-fold the emitted one (in red). The total unbalanced annual rate (in grey) has been used to drawn some function in the preceding Fig. 5. The water vapor is dissociated in the upper atmosphere by UV radiation and a percentual of Hydrogen reach the escape velocity. The consequent annual rate of atmospheric loss of water is estimated to be several order of magnitude less than the mantle's water rates (≈5·10⁻⁴ km³/y), and it is not represented in this diagram. Similarly, the possible contribution of water coming from space by small comets and small bodies of ice (≈3.6 ·10⁻⁴ km³/y) is too small to be represented.

In first approximation, assuming the mean depth of the oceans $h_{oce}$=const=4 km,

$$V_{oce} = (4\pi R^2 - S_{con}) \times 4.0 \text{ km},$$

which function is shown in Fig. 5.

The function is nearly linear and indicate an annual rate of volume increase of

$$\Delta V_{oce} \approx 5.6 \text{ km}^3/y, \quad (4)$$

without an average change of the sea level.

## Water Volume inputs or losses & their effects on sea-level

| | | Water Volume (km³/y) | Sea-level Change (cm/century) | Comment |
|---|---|---|---|---|
| i | Isostatic compensation of curvature bulges | | ≈ 1.5 – 15.0 | in the range useful to close sea-level eq. (2) |
| ii | Water displaced by erosion of bulges | 8.3 | ≈ 2.0 | upper limit |
| iii | Needed water input if all seafloors enlarge | 5.6 | 0.15 | upper limit if steady flow without expansion |
| iv | Hypothetic subducted water | 0.87 | $2.4 \cdot 10^{-2}$ | unpleasant and paradoxical: lead to Earth's desiccation |
| v | Estimated mantle water outpouring | 0.2 | $5.5 \cdot 10^{-3}$ | destined to be revised toward the higher value iii |
| vi | Water loss to outer space | $6.5 \cdot 10^{-4}$ | negligible | irrelevant in sea-level change problems |
| vii | Water incoming from space | $3.65 \cdot 10^{-4}$ | negligible | irrelevant in sea-level change problems |

Table 1. The variety of water volume inputs and losses and their effects on sea-level change. Preponderant effect – worthy to be compared with other effects (e.g. thermal expansion) accounted for in equation (2) – is expected to come from i), the isostatic adjustment of the higher curvature bulge, albeit 15 cm/century is an upper limit. A non negligible contribution come from ii), the uplift of the waters caused by erosion of the bulges. All the other contributions iii) to vii) are insignificant. The contribution v) in the expanding Earth conception needs a revision towards the higher value iii) because the expanding oceanic basins must be filled by juvenile water.

In the expanding Earth concept without subduction (Vogel, 1984; Cwojdziński, 2003; Maxlow, 2001; among others) or without large scale subduction (Scalera, 2010), a relevant reintroduction of water into the mantle is not allowed. The annual rate of water volume increase – $\Delta V_{oce} \approx 5.6$ km³/y, (eq (4)) – is more than 25 times the estimated annual rate of water outpouring in magmatism of arcs and ridges (Fig. 6). Consequently two not mutually excluding processes can be envisaged:

i) A greater rate of outpouring must be hypothesized. The ocean bottoms are still today largely unexplored and the possibility exists of a very large number of undetected spot-like or diffused sources of juvenile water. This hypothesis is strengthen by the interpretation of the trench-arc zones as regions of uplift of deep mantle materials, possibly hydrated, releasing water in a myriad of ocean bottom sources. Also the land sources of juvenile water should be reassessed in their real percent of content.

ii) A greater rate of tectonic activity in the geologic time lapses indicated in the *Half Spreading Map of the Oceans* (Müller et al., 1997; McElhinny & McFadden, 2000), where at least two periods of enhanced sea-floor expansion are present intercalated between the minima of Recent, Cretaceous-Cenozoic boundary and Jurassic-Cretaceous boundary (Fig. 2). A strongly enhanced emission of juvenile water should be expected in this more tectonically active periods.

The enlarging room of the oceanic basins should have evolved in average synchrony with the water mantle outpouring – or coming from other sources – (red line in Fig. 5), a synchrony process that may be has been favored by the isostatic adjustment of the ocean bottom depth (more emission leading to more high sea level and consequent isostatic depression of sea-bottom – the deep oceans has been created by the waters). We cannot admit that this synchrony is perfect along the geologic time axis, but it can be supposed that some phase shift may occur. For example beside a stasis of global expansion and tectonic activity (Fig. 2) two different hypothesis about the water mantle emission can be formulated: i) a parallel stasis of this emission, ii) a more or less intense prolongation of the emission. An upper limit can be then evaluated for the sea level change from equation (4), assuming a steady filling water flux of 5.6 km³/y in a situation of constant oceanic basins surface of $360.7 \times 10^6$ km²:

$\Delta h_{oce} \approx 0.15$ cm/century,

an the order of magnitude lesser than the values that can be taken into account to close equation (2). This means that the average water supply indicated by (4) is not sufficient to sustain a non-negligible portion of recent sea-level rise. Only hypothesizing a link between a *stasis* of the global expansion and an intensified orogenic activity (see the proposed orogenic model in: Scalera, 2007a, 2010, 2012d) an higher flow of juvenile mantle's water can be suspected. However, a two order of magnitude higher mantle water outpouring in order to get values relevant for equation (2) is hardly conceivable.

## 6 - Not closed water cycle: atmosphere water content

A subject never completely defined is the study of the rates of arrive, formation and loss of chemical species in the upper atmosphere. Water can be transported in this region from below or from the external space.

The comets have been suggested as an important supply of water to our planet (Frank et al., 1986; Lebedinets & Kurbanmuradov, 1992; Deming, 1999; among others). Time by time, this supply would be episodic, somewhat catastrophic (Shields, 1988), accompanied by sudden ingressions and slow regressions of the seas, or more continuous, powered by frequent fall of cosmic ice meteors (Bérczi & Lukács, 2001).

It has been evaluated that $10^3$ tons to more than $10^4$ tons – at least an order of magnitude of uncertainty (!) – of meteoric material falls on the Earth each day and that a substantial portion of it or a comparable additional amount could be constituted by icy bodies, which are more difficult to detect because they, with few exceptions, completely vaporize. If arbitrarily the upper value $10^4$ tons/day is assumed for the icy bodies, it can be derived a value of 0.000365 km$^3$/y, a small percentage of the mantle water supplies represented (red bars) in Fig. 6, that however must be considered a value possibly subject to future increasing re-evaluation.

For what concerns the possible loss of water to outer space, remember that the total content of water in the Earth's atmosphere is estimated to be only 0.001 % of the total water content of the hydrosphere. This amount is several order of magnitude greater than the amounts shown in Fig. 6. Then, an incorrect deduction would be that the loss for photodissociation and hydrogen escape out of the atmosphere must therefore be a very small and negligible percentage, because atmosphere is continuously replenished by water vapor from oceans (97.3 %), glaciers (2.1 %), lands (0.6 %), and biologic organisms (0.00004 %) (Lenz, 2013). In addition, some planets have lost their atmosphere in dependence of their particular condition of pressure and temperature. We only can hope that the percentage is small. Indeed, recent estimate points to a water loss of $5 \cdot 10^{11}$ grams lost each year by photodissociation (Brinkmann, 1969; Hunten & Donahue, 1976; Hunten et al., 1989; Lenz, 2013), which is about $6.5 \cdot 10^{-4}$ km$^3$/y, an amount several order of magnitude smaller than the amounts shown in Fig. 6, and which should be considered an upper limit because oxygen can recombine with hydrogen of different origin to build new water molecules.

It is possible to conclude that, with the present level of knowledge, we have not to care of the possible water supply from external space or loss of water by photodissociation. At least today, sea level is affected by these phenomena in a negligible manner, independently of any global tectonics theory.

## 7 - Different rotational effects

The consequences of the melting of the of polar ice sheets and of other glaciers has been considered by Munk (2002) in his concluding remarks:

> *The rotational evidence, although convoluted, appears to rule out a large eustatic contribution from melting on Antarctica and Greenland, assuming that the measured $j_2$ is representative of the 20th century. However, an enhanced contribution from glacial melting and other midlatitude sources is NOT ruled out by the rotational evidence.*

Albeit this is not the paper in which to try resolving in full details these discrepancies, it must be stressed that the effects of the combined global expansion and isostatic relaxation of the persistent curvature bulges could be very different in comparison to those caused by the melting of the ice caps. The ice melting produces a transferring of water from high latitudes towards intermediate and low latitudes. The flow of eroded materials from continents and the isostatic variations of the level of lands and ocean bottoms could lead to still unknown or not well quantified contributions to both spin-up and slow-down of the Earth.

These rotational effects should be considered worth to be investigated. Indeed, the impossibility to rule out enhanced contributions from glacial melting and other midlatitude sources (in the preceding quoted Munk's conclusion) is immediately evoking the scenery of erosive and isostatic phenomena involved in an expansion of the Earth.

## 8 - Discussion and conclusion

Water volume inputs, displacements and losses and their effects on sea-level change have been listed in Table 1. The exchange of water or its dissipation from higher atmosphere with outer space vi) and vii) – which can have importance in different phisical and thermal planetary conditions – are completely insignificant for sea-level change on actual Earth.

The quantity v) is the estimated emission of juvenile water from Earth's mantle, which could produce an upper limit of sea-level rise of only $5.5 \cdot 10^{-3}$ cm/century, in the case the same emission is maintained during a period of stasis of the globe expansion. However the emitted water volume of 0.2 km$^3$/y is too small to fill the progressively enlarging volume of the oceanic basins, iii), that amounts to 5.6 km$^3$/y.

A necessary condition for the validity of the expanding Earth concept is a revision towards higher values of the evaluated mantle outpouring, which must be more than 25 times greater. In any case this mantle emission does not influence in a substantial way the observed sea-level change.

The emission of fluids from Earth's interior can be linked to the researches mentioned in Section 5 (Bolfan-Casanova, 2005; Ohtani, 2005; Hirschmann, 2006; Hirschmann & Kohlstedt, 2012) that identify the regions of possible water storage, from which – reversing their interpretation of the path – it can be transported to the surface.

The future more complete exploration of the seafloors could lead to the discovery of a larger set of localized or diffused sources of juvenile mantle's water, or a reassessment of the juvenile water content of the existing sources.

Major effects on sea-level could come only from i) and ii) (Table 1), which are linked to ongoing curvature variations. The value i) of 15 cm/century is obviously an upper limit, but also a fraction of it, besides the value ii) coming from the erosive discharge to oceans of the bulges, can contribute to remove the difficulties highlighted by Munk (2002), Douglas & Peltier (2002), Miller & Douglas (2004).

The possibility to proof – at least as fitting order of magnitude – that in the expanding Earth framework it is possible to find additional phenomena that could contribute in a proper way to the water balance involved in the sea lever rising, should be considered a further support to the concept of global *emmitting Earth*, which is a concept that underlies the more general one of the expanding Earth.

This paper has no pretense to provide a detailed quantitative account of the effects caused by the curvature change on sea-level. Its aim are limited to scrutinize the involved magnitude orders and to compare them with the already recognized physical processes – like thermal expansion – that are cause of sea-level change.

The existence of unaccounted global geophysical processes linked to an expanding Earth can produce discrepancies in mainstream theories of sea level rise, and, as a matter of facts, this is the situation of the present day comparison between theory and observations. These discrepancies will presumably found a way toward their solution if new research lines will start that take into account the global isostatic and emissive phenomena of an expanding planet.

## Acknowledgements

Albeit I do not know him personally, I am greatly indebted to Walter Munk for having instilled in me the belief (a *meme* that has acted on my mind long after reading of his work in 2002) that the unsolved problems of the change of sea level would be discussed taking additionally into account an expansion of the globe.